# Charge Effects and Nanoparticle Pattern Formation in Electrohydrodynamic NanoDrip Printing of Colloids

Patrizia Richner, Stephan J.P. Kress, David J. Norris, Dimos Poulikakos*

P. Richner, Prof. D. Poulikakos
Laboratory for Thermodynamics in Emerging Technologies, ETH Zurich, Sonneggstrasse 3, 8092 Zurich, Switzerland
E-mail: dpoulikakos@ethz.ch

S. J. P. Kress, Prof. D. J. Norris
Optical Materials Engineering Laboratory, ETH Zurich, Leonhardstrasse 21, 8092 Zurich, Switzerland

**Abstract**

Advancing open atmosphere printing technologies to produce features in the nanoscale range has important and broad applications ranging from electronics, to photonics, plasmonics and biology.[1-4] Recently an electrohydrodynamic printing regime has been demonstrated in a rapid dripping mode (termed NanoDrip), where the ejected colloidal droplets from nozzles of diameters of O(1 µm) can controllably reach sizes an order of magnitude smaller than the nozzle and can generate planar and out-of-plane structures of similar sizes.[1] Despite demonstrated capabilities, our fundamental understanding of important aspects of the physics of NanoDrip printing needs further improvement. Here we address the topics of charge content and transport in NanoDrip printing. We employ quantum dot and gold nanoparticle dispersions in combination with a specially designed, auxiliary, asymmetric electric field, targeting the understanding of charge locality (particles vs. solvent) and particle distribution in the deposits as indicated by the dried nanoparticle patterns (footprints) on the substrate. We show that droplets of alternating charge can be spatially separated when applying an ac field to the nozzle. The nanoparticles within a droplet are distributed asymmetrically under the influence of the auxiliary lateral electric field, indicating that they are the main carriers. We also show that the ligand length of the nanoparticles in the colloid affects their mobility after deposition (in the sessile droplet state).

**Introduction**

Electrohydrodynamically driven liquid ejection from nozzles (EHD) has been known for several decades.[5-7] Different regimes of EHD ejection have been investigated and categorized,[8] the most well-known of which being the Taylor cone and jet regime, where a continuous jet emanates from the apex of a sharply focused meniscus at the opening of a nozzle. Due to instabilities, this jet may break up into droplets containing like charges, which are then sprayed onto a surface.[4, 9] In the so-called microdripping regime, the applied electric field is smaller, the meniscus less focused and instead of a jet, a small portion of the meniscus at its apex breaks off continuously, generating a continuous stream of droplets markedly smaller than the size of the meniscus itself and the nozzle opening. Recent studies have realized a dripping regime at the nanoscale (NanoDrip), able to generate structures from colloidal inks not only with ultra-high planar resolution, but also out-of-plane[1, 3] and on non-flat, prestructured substrates.[2, 10]

In the NanoDrip process, typically, a gold-coated nozzle with an opening diameter of 1-2 µm filled with a nanoparticle dispersion is brought within a few (1-10) µm of the substrate. When applying an electric field between the nozzle and the substrate, droplets with diameters down to a good order of magnitude smaller than the nozzle opening are ejected. Each droplet lands on the substrate and the solvent evaporates before the arrival of the subsequent droplet, leaving behind only the pattern formed by the dried nanoparticle content of the ink.

This versatile technology can be used in a broad range of applications, such as for cancer cell studies,[3] printed electronics,[11-13] plasmonic applications[2, 10] or single crystal deposition.[14] The versatility is accompanied by a host of related challenges when it comes to substrates (soft and hard, conductive and nonconductive, flat and prestructured) and the materials to be deposited in terms of ink content (metal or semiconductor nanoparticles and metal salts). In order to define the optimal working conditions for each application, a thorough understanding of the physical processes involved in the ejection and deposition of the nanoparticle dispersions is needed. Based on these findings the printing process parameters and the nanoparticle dispersions could then be tailored according to the needs of a specific application.

Several studies have investigated the properties of the EHD spraying in the Taylor cone regime for a range of liquids, such as glycerin or distilled water.[15-17] However, despite progress and their obvious importance, our knowledge of charge transport mechanisms in the nanodripping regime is still incomplete. Here we show how the charge transport for EHD NanoDrip printed nanoparticle dispersions differs fundamentally from conventional EHD spraying. In the Taylor regime and for more conductive liquids, naturally existing ions in the liquid ink (e.g. by dissociative processes, $H_3O^+$ and $OH^-$ in the case of water) or molecules easily ionized (for example because they have double bond structures) will act as charge carriers. Depending on the polarity of the applied voltage, electrons or holes are transferred to the liquid. On the other hand, if the goal is to precisely print nanoparticle structures with a submicron spatial resolution, as often the case in the NanoDrip process, the choice of solvent is constrained by other factors necessary for the process. For example, the solvent needs to combine with the various nanoparticles to form stable dispersions for different applications while not containing any significant amount of free ligands or other byproducts as these may adversely influence the properties of the printed material. In addition, the vapor pressure employed has an upper bound, because fast evaporation may clog the small nozzle (order of a micron nozzle opening) as well as a lower bound, since the evaporation of a droplet on the substrate must be complete before the arrival of the next droplet, to avoid puddle formation, which would be detrimental to the generation of nanostructures and to out-of plane nanoprinting.[1, 3, 18]

In this work we investigate charge transport in the nanodripping regime by printing nanoparticles dispersed in tetradecane, a solvent fulfilling all of the above-mentioned requirements. The electrical conductivity of tetradecane in its pure form was measured to be as low as $3 \cdot 10^{-12} S/cm$, which is 4 orders of magnitude lower than ultra-pure water. Ions by dissociation, or molecules with double bonds do not exist in tetradecane and can hence not act as charge carriers. Electrohydrodynamic ejection of pure tetradecane in a dripping mode is consequently not possible (we have shown this experimentally in earlier works).[19, 20] Hence, unlike in the Taylor cone process, it is highly improbable that the electric charge transport stems from the solvent itself. The aim is to gain an in-depth understanding of the charge content in droplets ejected by NanoDrip printing and its effect on the forces acting on the droplets and the nanoparticles in the droplets. To this end, we show experimentally that the charges in ejected droplets in the nanodripping regime are carried by the nanoparticles themselves rather than the solvent. Whether the charges are distributed throughout a liquid solvent, under the prerequisite that the liquid is conductive enough to carry the appropriate amount of charges, or whether they are localized on nanoparticles, does not fundamentally change the ejection mechanism. When the nanoparticles carry the charges, they will move to the surface of the meniscus and then induce electrohydrodynamic dripping (the specific mode of our work), just like ions do in the case of more conductive liquids. Since we cannot observe the droplet flight directly due to the extremely small length and time scales involved, we modulate the droplet orbit while in flight with a specially designed asymmetric electric field and draw conclusions from the study of the patterns of the dried nanoparticle content on the substrate (termed footprints) after solvent vaporization, a posteriori.

**Materials and Methods**

All experiments are carried out on an in-house-built printing setup (Figure 1a). A glass capillary is pulled into a nozzle with a Sutter P-97 pipette puller. The nozzle, with an opening diameter of approximately 2 µm, is then coated with 10 nm Ti and 100 nm Au and rendered conductive. The substrate is mounted on a 3D piezo-stage from MadCityLabs, the nozzle is filled with the nanoparticle dispersion (self-driven by capillarity) and is brought within 10 µm of the substrate. In the standard printing mode, the sample is positioned on a grounded plate, and ejection is induced by applying a voltage of 125-250 V to the nozzle. In order to study the charge transport of electrohydrodynamic ejection in the nanodripping mode, we added a lateral component to the electric field, realized with the help of gold electrodes designed for this purpose, with a separation of 2 µm deposited on a glass wafer with conventional photolithography. This allowed us to apply an electric field between these lateral electrodes. A scanning electron micrograph (SEM) image of the lateral electrodes before nanoparticle deposition is shown in Figure 1b, where the ground and the positive electrode are marked. When printing in a straight line along the x-axis between the lateral electrodes, the ejected, charge carrying droplets are deflected in the y direction, normal to the direction of motion of the stage. The deflection of the droplets is a function of the voltage applied to the lateral electrode. The lateral voltage has to be low enough such as not to interfere with the electric field at the nozzle, as this would disturb the ejection process and yet high enough to yield a measurable deflection on the substrate. We employ a constant lateral voltage of 30 V throughout this study which fulfills both conditions. In order to study the properties of a single droplet footprint, the stage is moved rapidly to ensure sufficient separation of the footprints of individual droplets.

Three useful nanoparticle dispersions covering a range of applications[2, 10, 21] are used in this work. These are CdSe-CdS-ZnS core-shell-shell quantum dots with a diameter of 10 nm coated with oleic acid and 5 nm gold nanoparticles stabilized with either octanethiol or dodecanethiol. All three inks have been described in detail in earlier publications.[9,10, 21] In order to study the polarity of the droplets as well as the charged nanoparticles we used the quantum dot ink, because of its better visibility in the SEM due to the larger size of the quantum dots compared to the gold nanoparticles.

**Results and Discussion**

In Figure 2 three SEM images show footprints transitioning, going from left to right, from absence to presence of a lateral electric field. The ground and positive lateral electrodes are labeled and the footprints are artificially colored according to their charge (red: positive, blue: negative) for better visibility. The dashed line is a guide to the eye to mark where the droplets would land in the absence of a lateral field. If a positive or negative dc pulse is applied to the nozzle in the NanoDrip mode, charged droplets are ejected towards the substrate as can be seen in Figures 2a and 2b, respectively. The droplets are deflected towards the lateral ground or positive electrode when printing between the electrodes (right-hand side of the picture). No deflection is seen on the left hand side, where there is no influence of the lateral electric field and the footprints show the typical impact pattern of submicron quantum dot dispersion droplets.[22] In Figure 2c an ac voltage is applied between the nozzle and the ground plate. Since the colloidal ink is expected to contain both positive and negative charges with no apparent preference, this leads to the ejection of positively charged drops during the positive phase of the signal and, vice versa, negatively charged drops during the negative phase of the signal. Accordingly, a separation of positively and negatively charged droplets between the lateral electrodes can be observed when the lateral electric field is present, while all droplets land on a straight line without the influence of this field (colored in green).

In Figure 3 close-up SEM micrographs of quantum dot footprints are shown to demonstrate the effect of the lateral electric field. Figure 3a shows a footprint outside the region of the lateral electrodes for comparison purposes. There is a clear rotational symmetry in the deposition of the quantum dots. Due to the relatively large size of the droplet there is an outward flux of the nanoparticles, leading to a partial coffee-stain effect.[23] Figures 3b-d show footprints placed between the lateral electrodes. To this end, Figures 3b and c show footprints from a positively and a negatively charged droplet, respectively, and Figure 3c the footprints of droplets ejected by an ac signal.

The shape of the footprints is now oval with the nanoparticles showing a clear tendency to concentrate towards the direction of the electric field, indicated by the arrow and violet shading in Figure 3b. The electric field on the substrate between the lateral electrodes is uniform, such that a dielectrophoretic force (proportional to field gradients) can be excluded. The evaporation of the solvent is rotationally symmetric and hence evaporative fluxes cannot be responsible for such an asymmetric placement of the nanoparticles. The explanation is that nanoparticles, themselves electrically charged, move in the electric field until they reach the drop boundary (contact line region). The interplay of the coffee-stain effect and the lateral electric field then leads to the particular oval footprint shape shown in Figures 3b-d. The total number of quantum dots in the footprint in Figure 3b is 1360. If they were distributed symmetrically, the violet area would contain about 470 particles, while in reality there are 750 particles. The difference of roughly 280 particles carry charge and are therefore pulled to one side by the electric field.

The claim that the charges are located on the nanoparticles rather than in the solvent can be further underpinned by examining the Born energy of an ion:[24]

$$\mu = \frac{Q^2}{8\pi\varepsilon_0\varepsilon a} \qquad (1)$$

Where $\mu$ is the Born energy, Q the charge of the ion, a is the radius of the ion, $\varepsilon_0$ and $\varepsilon$ are the vacuum permittivity and the dielectric constant of the medium, respectively. The energy can be minimized by moving the charge from a medium of low dielectric constant to a medium of higher dielectric constant. In the case of the quantum dot dispersion, CdSe has a higher dielectric constant than tetradecane and is therefore a favorable location for the charges. The energy minimization holds even more for metallic nanoparticles, where a charge can be delocalized over the volume of the nanoparticles. The ejection frequencies observed in the experiments here indicate that it takes about 10 ms to eject a single droplet. The thermodynamic energy minimization process is of a much shorter time scale, hence the charges are located on the nanoparticles already at the onset of the droplet ejection.

Knowing the weight of the droplet, the nozzle-substrate distance and the magnitude of the involved electric fields it is possible to estimate the charge carried by one single droplet by measuring the deflection of the footprints between the electrodes. It has been shown earlier[1] that in NanoDrip printing the diameter of the footprint is of the same order of magnitude as the diameter of the corresponding airborne droplet. We can hence estimate the weight of one droplet generating any given footprint. The nozzle-substrate distance is controlled with the piezo-stage and the electric field in the space between the nozzle and the substrate can be determined with numerical simulations.

The forces acting on the flying droplet are threefold: The force due to the electric field in the downward z-direction, the force due to the electric field in the lateral direction induced by the side electrodes and the Stokes drag for laminar flow around each droplet in flight, directed against the flight direction.

We carried out 3D electrostatic simulations with the commercial software COMSOL. A color plot of the cross-section of the electric field in the region between the nozzle and the substrate with the lateral electrodes is depicted in Figure 4a. The hemisphere at the tip of the nozzle represents the liquid meniscus, which is generated when a voltage is applied to the nozzle (the apex of this meniscus is what is printed in the NanoDrip mode, as depicted in Figure 1a). COMSOL offers a feature where the flight path of a charged particle (here droplet) with predefined mass and charge is calculated. It also allows accounting for additional forces such as the Stokes drag, which for spheres of size comparable to the mean free path in air is given by:

$$F_D = -\frac{6\pi\eta R v}{1+Kn\left(a+be^{-\frac{c}{Kn}}\right)} \quad (2)$$

where $\eta = 1.983 \cdot 10^{-5} Pa s$ is the viscosity of air, R the droplet radius, v the droplet velocity, $Kn = \frac{\lambda}{R}$ the Knudsen number with λ the mean free path of air, and a=1.252, b=0.399 and c=1.1 are empirical coefficients to account for the Cunningham slip correction factor for droplet sizes comparable to the mean free path of air.[25]

In Figure 4b the flight paths from the nozzle to the substrate for droplets of equal size but varying charges are shown. The deflection decreases with increasing charge. The greater the charge of a droplet is, the stronger its acceleration after leaving the nozzle and the higher its velocity when entering the space where the lateral electric field is appreciable. This has a negative effect on the droplet lateral deflection before landing on the substrate. The deflection of a positively charged droplet ejected from a nozzle with a potential of 200 V is shown in Figure 2a and measured to be 450 nm. For an order of magnitude estimate we assume the droplet diameter in-flight to be about 200 nm.

The Rayleigh limit poses a natural upper bound to the charge $Q_0$ a liquid drop can carry:

$$Q_0 = 8\pi\sqrt{\varepsilon_0 \gamma R^3} \qquad (3)$$

where $\varepsilon_0$ is the vacuum permittivity, $\gamma$ the surface tension of the droplet (for tetradecane[26] γ=26.56 mN/m) and R is the droplet radius. The maximum charge a droplet with a diameter of 200 nm could carry before breaking up is hence estimated to be $3.9 \cdot 10^{-16}$ C, which corresponds to about 2400 elementary charges. The deflection of the droplets in the experiment is indicated with a dashed line in Figure 4b, and corresponds to a charge of $3.6 \cdot 10^{-17}$ C or about 225 elementary charges, which is close to 10% of the Rayleigh limit. Other groups have measured the charge of EHD ejected droplets in a larger scale dripping mode (generated droplet diameter in the range of a few µm to a few hundred µm), and found charges ranging from around 10% of the Rayleigh limit[27] up to about half the Rayleigh limit[8]. While the large charges were measured for droplet diameters of 200-300 µm, the small charges were measured for droplets in the range of a few µm. Since the droplet diameters in this work are even smaller, we believe our results are in acceptable agreement with the existing literature. Since we estimate the inflight droplet diameter based on the footprint diameter, to quantify the effect of this estimate, the simulated flight paths for droplets with a range of diameters (150-300 nm) and a charge corresponding to the same percentage of the Rayleigh limit (eq. 3) as the 200 nm droplet in Figure 4b is plotted in Figure 4c. The larger a droplet is, the smaller its deflection due to its greater inertia and the higher its speed close to the lateral electrodes. Hence the 150 nm droplet sets the upper bound for the depicted range and the 300 nm droplet the lower bound.

When comparing Figures 3a and b we see that a considerable population of nanoparticles is displaced in the presence of the lateral field. While it is not possible to directly measure the individual quantum dot charges, the roughly 280 asymmetrically displaced quantum dots in Figure 3b are of the same order of magnitude as the number of elementary charges per droplet predicted by the simulations. The same effect can be observed in Figure 3c, where negatively charged quantum dots moved upwards toward the 30 V electrode and in Figure 3d, (ac-field at the nozzle) where droplets of alternating charge are ejected and are deflected alternatingly upwards and downwards according to their respective charge.

The above evidence shows that the droplets as whole carry either positive or negative charges and that the nanoparticles themselves are the carriers of these charges. The assessment of the forces acting on the nanoparticles requires a more detailed discussion. The main forces acting between the substrate and the nanoparticles in a sessile droplet are van der Waals-forces and – if the nanoparticle carries a charge – Coulomb forces. Given the electric field, the Coulomb force can be calculated by multiplying this field with the charge. The van der Waals-forces are caused by electromagnetic interactions between the nanoparticle and its surrounding, comprising not only the glass substrate and tetradecane but also other nanoparticles and the surfactant on the nanoparticle. We circumvent the obvious complexity in estimating the magnitude of the force acting on a nanoparticles in solution close to the substrate, by drawing conclusions from the simple case of the dependence of such a force on the distance from the substrate considering the arrangement of a sphere and an infinite plane separated by a small distance in vacuum. The force is then given by:[24]

$$F(d) = -\frac{AR}{6d^2} \qquad (4)$$

Where F is the force acting on a single nanoparticle, A is the Hamaker constant accounting for the two interacting materials, here gold and glass, R is the nanoparticle radius and d the distance between the particle surface and the substrate. In order to exclude all other influences, we compare two gold nanoparticle dispersions, which differ from each other only by the length of the octanethiol and dodecanethiol ligand. The length of a thiol-capped surfactant on gold nanoparticles with a diameter of 5 nm has been measured to be 0.73 nm for octanethiol and 0.93 nm for dodecanethiol by Wan et al.[28]. These lengths define the smallest distance between the nanoparticle and the substrate. Considering that the van der Waals-force is inversely proportional to the square of the separation (eq. 4), the force magnitude for dodecanethiol-capped nanoparticles is 60% of the force magnitude for octanethiol-capped nanoparticles. The effect of this difference on the strength of the van der Waals-force is illustrated in Figure 5, where two footprints located between the lateral electric field electrodes are shown, printed at a nozzle voltage of -150 V. The footprint in Figure 5a consists of octanethiol-capped gold nanoparticles and the one in Figure 5b of dodecanethiol-capped particles. While in the case of octanethiol there is no visible influence of the lateral field breaking the circular symmetry of the footprint pattern (Fig. 5a), a clear movement of the negatively charged particles towards the 30 V electrode can be seen for dodecanethiol-capped particles (Fig. 5b). While the Coulomb force is the same in both cases, the van der Waals-force is much stronger for the shorter surfactants, effectively inhibiting any particle movement.

**Conclusion**

In conclusion, we studied the physics of charge content of submicron-sized droplet colloids in tetradecane (effectively a dielectric liquid), electrohydrodynamically ejected in the nanodripping mode, by examining the deposited dry nanoparticle patterns (footprints). In addition to the vertical electric field for ejecting the droplets from the nozzle, we employed an auxiliary lateral electric field directly on the substrate, deflecting on the one hand the droplets in flight and on the other hand acting on the charges within the droplet after landing during drying. By combining experiment and simulations we estimated that the ejected droplets carry electrical charges of the order of 10% of the Rayleigh charge limit, in accordance with publications from other groups. The results strongly support the argument that such charges are carried by the nanoparticles, for both materials employed, quantum dots and gold nanoparticles. By comparing estimates of van der Waals-forces acting on a nanoparticle on the substrate, we further showed that depending on the length of the surfactant, the charged nanoparticles can be either immobilized or move in the applied lateral electric field.

**Acknowledgements**

We gratefully acknowledge fruitful discussions with J. Schneider and P. Galliker of LTNT as well as assistance from D. Kim and A. Riedinger of OMEL and P. Rohner of LTNT. The research leading to these results has received funding from the Swiss National Science Foundation under Grant 146180. We also gratefully acknowledge funding from the European Research Council under the European

Union's Seventh Framework Programme (FP/2007-2013)/ERC Grant Agreement Nr. 339905 (QuaDoPS Advanced Grant).

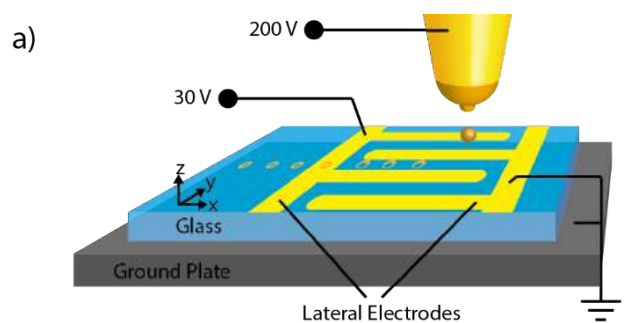

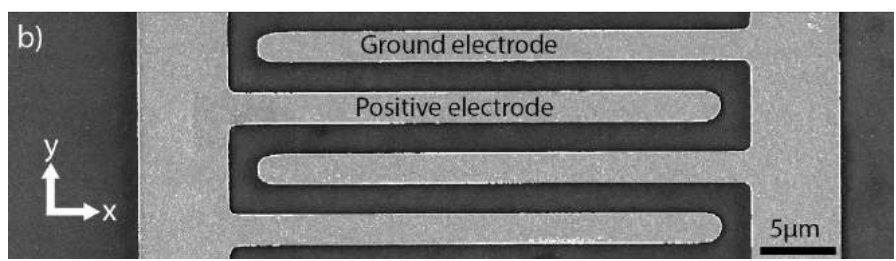

**Figure 1.** a) Schematic of the printing setup. The gold-coated nozzle in the z direction is kept at 200V at 10 µm from the glass substrate, while the substrate is moved in the x direction, parallel to the electrodes. Single droplets are deposited on the substrate, well separated from each other; b) Top view scanning electron micrograph of the lateral electrodes on the glass substrates: Droplets are printed in the gap between a ground and a positive electrode, leading to well-separated dried nanoparticle footprints.

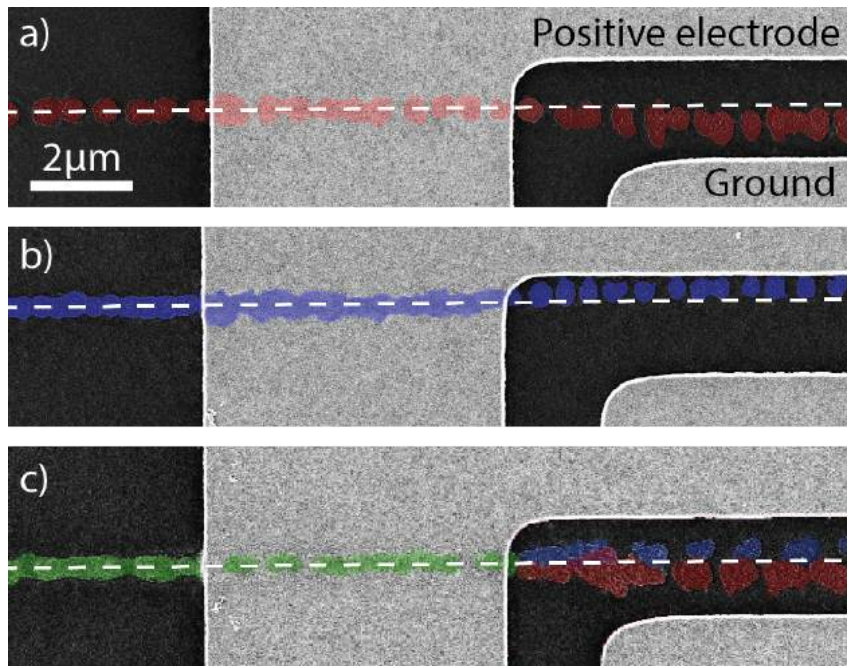

**Figure 2.** Scanning electron micrograph showing footprints transitioning from absence of a lateral electric field on the left to presence of a lateral field on the right; the positive and ground electrodes are labeled. The footprints are artificially colored according to their charge (red: positive, blue: negative, green: alternating) for better visibility. The dashed line is a guide to the eye showing where the footprints would be located in an undisturbed field (left side of the images). Results are shown for a positive nozzle voltage (a), a negative nozzle voltage (b) and an ac voltage at the nozzle (c). The droplets are deflected in flight, positively charged droplets towards the ground electrode and negatively charged droplets towards the positive electrode. In the case of an ac field at the nozzle ejection alternates between positively and negatively charged droplets.

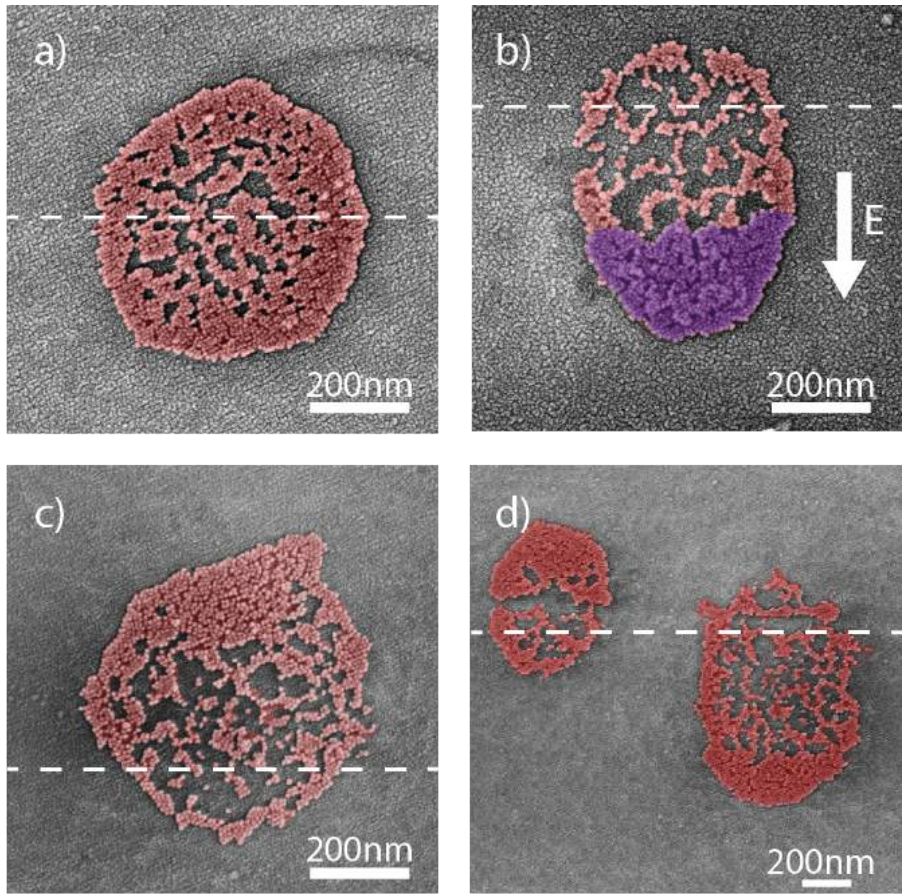

**Figure 3:** High magnification micrographs of footprints without lateral electric field (a) and in the presence of a lateral electric field (b-d) for positive, negative and ac voltages at the nozzle, respectively, the dashed line indicates where the droplet would land in an undisturbed field (absence of a lateral field). Comparing (a) and (b), a considerable population of quantum dots has moved downwards in the electric field (arrow) direction. The number of displaced quantum dots is around 280, which is of the same order of magnitude as the number of elementary charges per droplet as calculated in Figure 4.

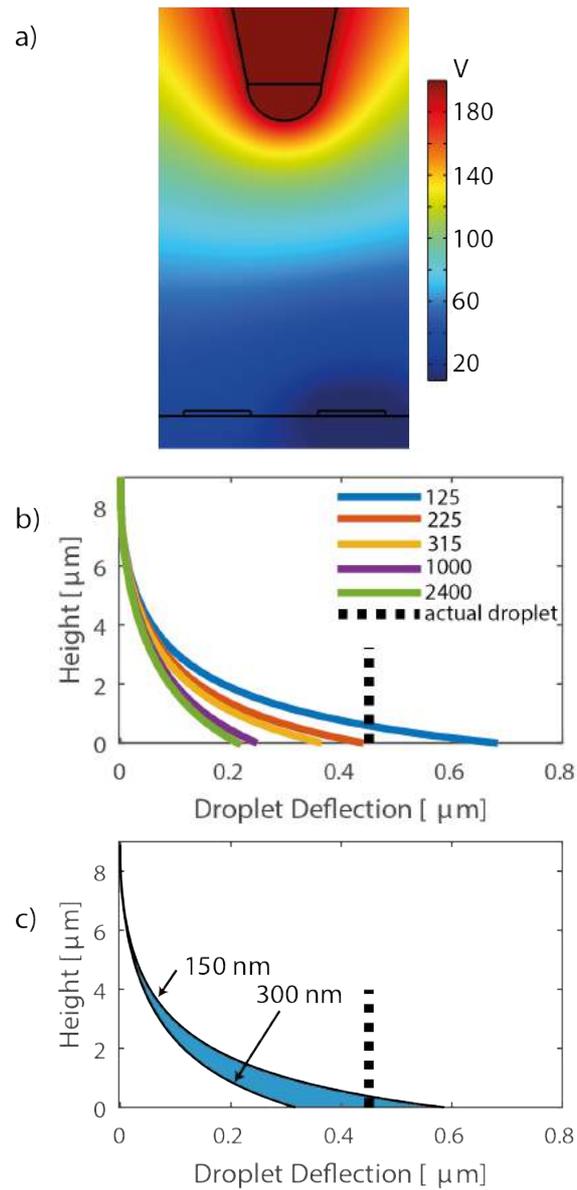

**Figure 4.** a) Cross section plot of the electric potential through the nozzle and the two lateral electrodes, held at 30 V on the left and grounded on the right. The hemisphere at the tip of the nozzle is tetradecane pulled out by the electric field. The unit is V; b) Droplet flight path simulated with COMSOL. The diameter of the droplet is 200 nm, the voltage at the nozzle 200 V, and the lateral electrodes at ground and at 30 V. Different flight paths as a function of the number of unit charges in the droplet are plotted. The dashed line indicates the deflection of 450 nm, as measured in Figure 2a. As can be seen, this deflection corresponds to about 225 unit charges; c) Droplet flight path for a range of droplet diameters of 150-300 nm, each containing roughly 10% of the charge of their respective Rayleigh limit.

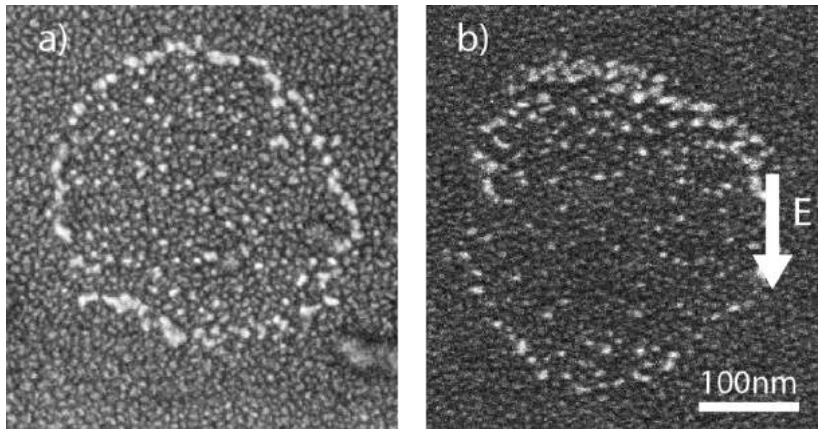

**Figure 5.** SEM micrographs of two gold nanoparticle footprints printed with -150 V at the nozzle, both footprints located between the lateral electrodes, the direction of the electric field is indicated. The surfactant on the nanoparticles is octanethiol (a) and dodecanethiol (b). Due to the weaker van-der Waals-forces the nanoparticles coated with dodecanethiol can move on the substrate while the ones with shorter surfactants are immobilized on the substrate.

# Charge Effects and Nanoparticle Pattern Formation in Electrohydrodynamic NanoDrip Printing of Colloids

Patrizia Richner, Stephan J.P. Kress, David J. Norris, Dimos Poulikakos*

P. Richner, Prof. D. Poulikakos
Laboratory for Thermodynamics in Emerging Technologies, ETH Zurich, Sonneggstrasse 3, 8092 Zurich, Switzerland
E-mail: dpoulikakos@ethz.ch

S. J. P. Kress, Prof. D. J. Norris
Optical Materials Engineering Laboratory, ETH Zurich, Leonhardstrasse 21, 8092 Zurich, Switzerland

**Deformation of the stationary droplet between the lateral electrodes**

The following simple model assesses the deformation of a charged, hemispherical droplet in the electric field between the lateral electrodes used in the main text. :

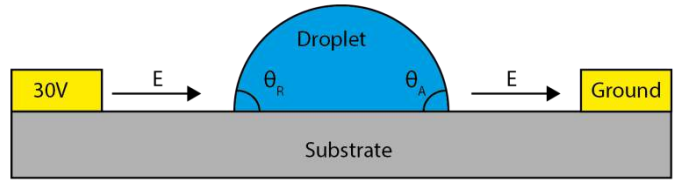

**Supplementary Figure 1:** Sessile droplet between the lateral electrodes, with receding and advancing angle indicated.

Following Gent's model[1] of a volumetric force acting on a droplet we obtain:

$$F_{el} = F_{hysteresis}$$

$$QE = \frac{4}{\pi} R \, \gamma_{C14}(\cos(\theta_R) - \cos(\theta_A))$$

Where Q is the total charge in the droplet, E the electric field, R the droplet radius and $\gamma_{C14}$ the surface tension of tetradecane. Further assuming a hemispherical droplet with a contact angle of 90° as a first approximation,[2] we take $\theta_R$=90°+δ and $\theta_A$=90°-δ for the receding and the advancing angle in the deformed droplet, respectively. For small δ the cosine function can be simplified and the equation can easily be solved for the deformation angle δ:

$$\delta = \frac{QE\pi}{2R\gamma_{C14}} = 1.8°$$

With Q=225$q_0$, E=1.5*$10^7$ V/m (homogenous field between electrodes at 30V, separated by 2µm), R=250nm (droplet contact radius, taken from Figure 3b in the main text), $\gamma_{C14}$=26.56mN/m[3]. This would lead to advancing and receding contact angles of 91.8° and 88.2°, respectively. This small deflection hardly justifies the massively asymmetric nanoparticle deposition we observe.

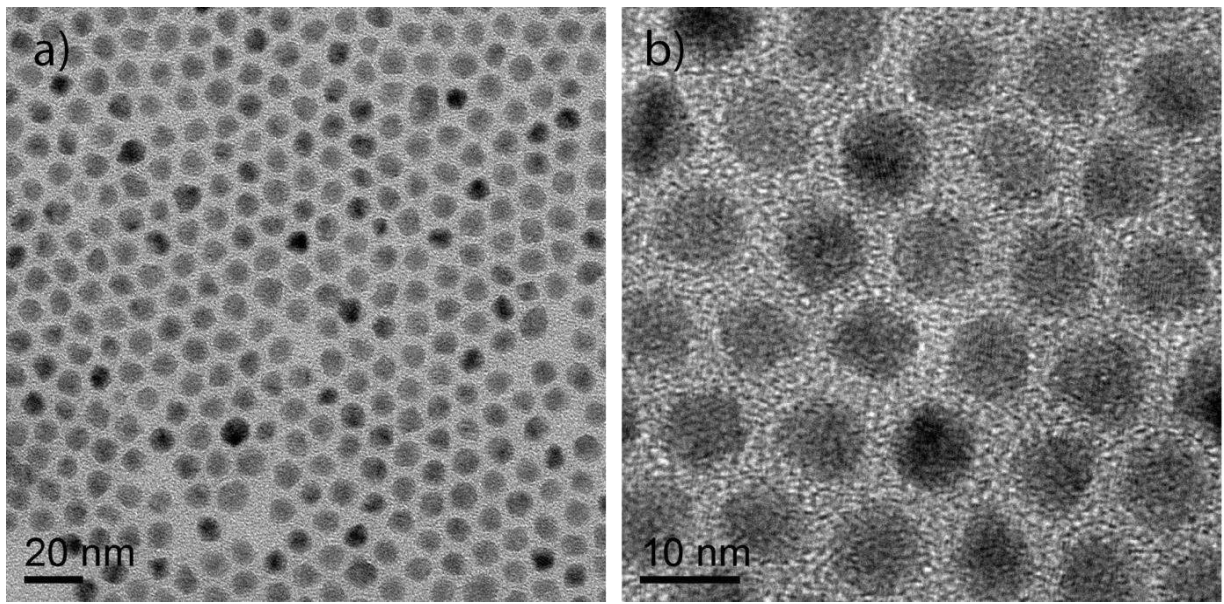

**Supplementary Figure 2:** TEM images of the quantum dots used for printing in the main text.

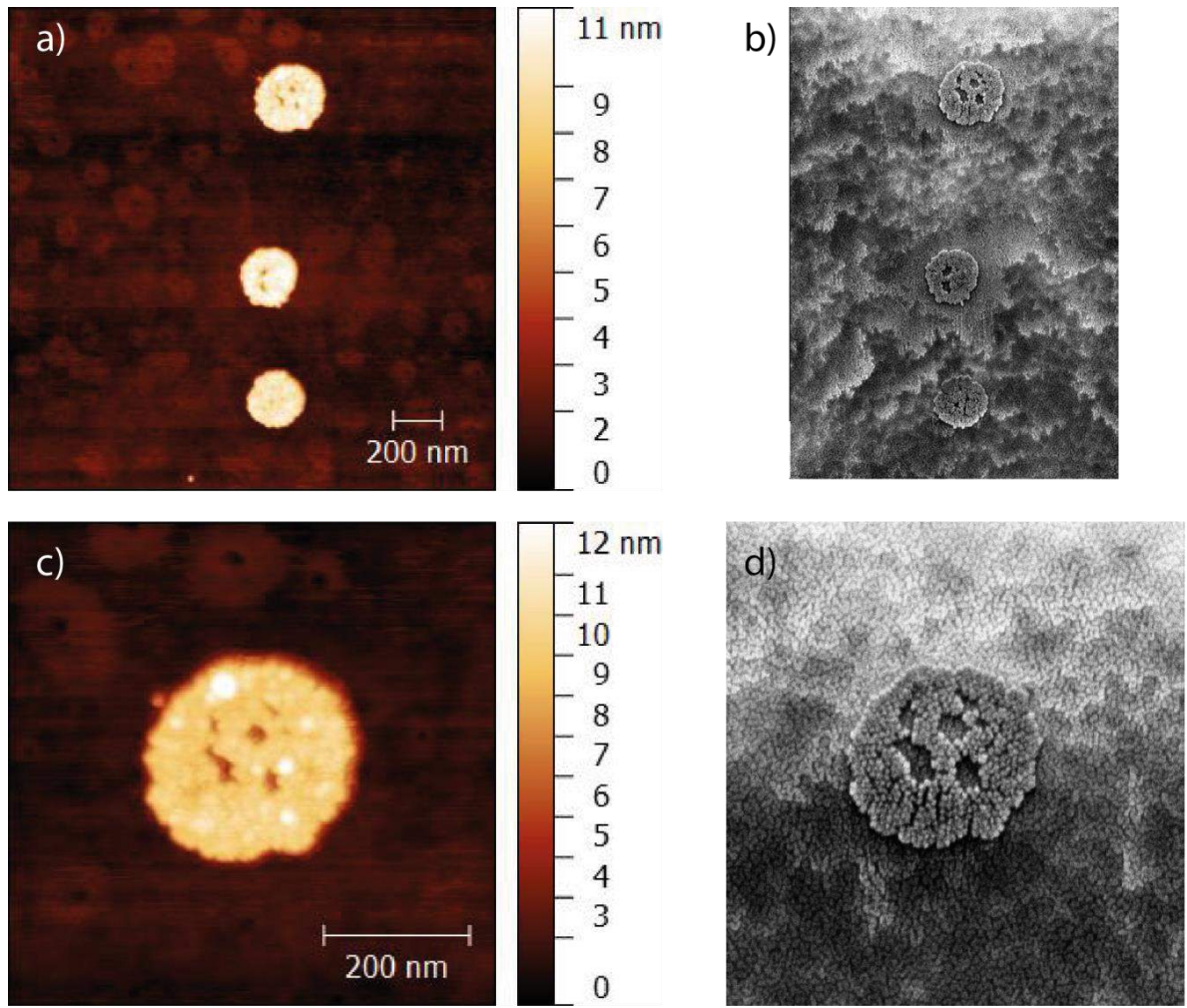

**Supplementary Figure 3:** AFM (a,c) and SEM (b,d) images of printed quantum dot footprints. The height of the footprints corresponds to one monolayer of quantum dots with a diameter of around 10 nm, as measured in the TEM scans. The AFM tip radius is nominally 8 nm, which means that the crevices between the single dots cannot be resolved. They can however very well be distinguished in the SEM graphs of the identical footprints on the right.